\documentclass[final,letterpaper,5p]{elsarticle}
\usepackage{lineno}
\usepackage{graphicx,color}
\usepackage{amssymb}
\usepackage{amsmath}
\usepackage{xspace}                                                 
\usepackage{latexsym}
\usepackage{amsxtra}
\usepackage{amscd}
\usepackage{dcolumn}
\usepackage{bm}
\usepackage{hyperref}
\usepackage{url}
\usepackage[T1]{fontenc} 
\usepackage{threeparttable}
\biboptions{numbers,sort&compress}
\graphicspath{{./}}

\newcommand{\nuc}[2]{$^{#1}$#2} 

\newcommand{\efm}{$e^2$fm$^4$\xspace}

\newcommand{\grays}{$\gamma$ rays\xspace}
\newcommand{\ghray}{$\gamma$-ray\xspace}

\newcommand{\pppf}[4]{$^{#1}$#2$(p,2p)$$^{#3}$#4}
\newcommand{\pppfr}[4]{$^{#1}$#2$(p,2p)$$^{#3}$#4 reaction\xspace}

\newcommand{\ppnf}[4]{$^{#1}$#2$(p,pn)$$^{#3}$#4}
\newcommand{\ppnfr}[4]{$^{#1}$#2$(p,pn)$$^{#3}$#4 reaction\xspace}

\newcommand{\mcnt}[1]{\multicolumn{2}{c}{#1}}

\newcommand{\fn}{\tnote}

\journal{Physics Letters B}

\begin{document}
\begin{frontmatter}
  
\title{A First Glimpse at the Shell Structure beyond \nuc{54}{Ca}: \\Spectroscopy of \nuc{55}{K}, \nuc{55}{Ca}, and \nuc{57}{Ca}}
\author[ut,rnc]{T.~Koiwai}
\author[ut,rnc,csic,gsi]{K.~Wimmer\corref{cor1}}
\cortext[cor1]{Corresponding author}
\ead{k.wimmer@gsi.de}

\author[rnc]{P.~Doornenbal}
\author[tudarmstadt,cea,rnc]{A.~Obertelli}

\author[miuni,miinfn,surrey]{C.~Barbieri}
\author[cea,leuven]{T.~Duguet}
\author[triumf,mcgill]{J.~D.~Holt}
\author[triumf]{T.~Miyagi}
\author[triumf]{P.~Navr\'atil}
\author[rcnp,ocu,nitep]{K.~Ogata}
\author[cns]{N.~Shimizu}
\author[cea]{V.~Som\`a}
\author[jaea,cns]{Y.~Utsuno}
\author[jaea]{K.~Yoshida}

\author[caen]{N.~L.~Achouri}
\author[rnc]{H.~Baba}
\author[rnc]{F.~Browne}
\author[cea]{D.~Calvet}
\author[cea]{F.~Ch\^ateau}
\author[beijing,rnc]{S.~Chen}
\author[rnc]{N.~Chiga}
\author[cea]{A.~Corsi}
\author[rnc]{M.~L.~Cort\'es} 
\author[cea]{A.~Delbart}
\author[cea]{J-M.~Gheller}
\author[cea]{A.~Giganon}
\author[cea]{A.~Gillibert}
\author[cea]{C.~Hilaire}
\author[rnc]{T.~Isobe}
\author[tohoku]{T.~Kobayashi}
\author[rnc,cns]{Y.~Kubota}
\author[cea]{V.~Lapoux}
\author[cea,kth]{H.~N.~Liu}
\author[rnc]{T.~Motobayashi}
\author[ipno,rnc]{I.~Murray}
\author[rnc]{H.~Otsu}
\author[rnc]{V.~Panin}
\author[cea,lkb]{N.~Paul}
\author[unal,rnc]{W.~Rodriguez}
\author[rnc]{H.~Sakurai}
\author[rnc]{M.~Sasano}
\author[rnc]{D.~Steppenbeck}
\author[cns]{L.~Stuhl}
\author[cea]{Y.~L.~Sun}
\author[rikkyo]{Y.~Togano}
\author[rnc]{T.~Uesaka}
\author[rnc]{K.~Yoneda}

\author[kth]{O.~Aktas}
\author[tudarmstadt]{T.~Aumann}
\author[inst]{L.~X.~Chung}
\author[ipno,caen]{F.~Flavigny}
\author[ipno]{S.~Franchoo}
\author[zagreb,tudarmstadt,rnc]{I.~Gasparic}
\author[koeln]{R.-B.~Gerst}
\author[caen]{J.~Gibelin}
\author[ewha,ibs]{K.~I.~Hahn}
\author[ewha,ibs]{D.~Kim}
\author[titech]{Y.~Kondo}
\author[tudarmstadt,gsi]{P.~Koseoglou}
\author[hku]{J.~Lee}
\author[tudarmstadt]{C.~Lehr}
\author[inst]{B.~D.~Linh}
\author[hku]{T.~Lokotko}
\author[ipno]{M.~MacCormick}
\author[koeln]{K.~Moschner}
\author[titech]{T.~Nakamura}
\author[ewha,ibs]{S.~Y.~Park}
\author[gsi]{D.~Rossi}
\author[oslo]{E.~Sahin}
\author[tudarmstadt]{P-A.~S\"oderstr\"om}
\author[atomki]{D.~Sohler}
\author[titech]{S.~Takeuchi}
\author[tudarmstadt]{H.~Toernqvist}
\author[csic]{V.~Vaquero}
\author[tudarmstadt]{V.~Wagner}
\author[lanzhou]{S.~Wang}
\author[tudarmstadt]{V.~Werner}
\author[hku]{X.~Xu}
\author[titech]{H.~Yamada}
\author[lanzhou]{D.~Yan}
\author[rnc]{Z.~Yang}
\author[titech]{M.~Yasuda}
\author[tudarmstadt]{L.~Zanetti}

\address[ut]{Department of Physics, The University of Tokyo, 7-3-1 Hongo, Bunkyo-ku, Tokyo 113-0033, Japan}
\address[rnc]{RIKEN Nishina Center, 2-1 Hirosawa, Wako, Saitama 351-0198, Japan}
\address[csic]{Instituto de Estructura de la Materia, CSIC, E-28006 Madrid, Spain}
\address[gsi]{GSI Helmholtzzentrum f\"ur Schwerionenforschung GmbH, Planckstr. 1, D-64291 Darmstadt, Germany}
\address[tudarmstadt]{Institut f\"ur Kernphysik, Technische Universit\"at Darmstadt, D-64289 Darmstadt, Germany}
\address[cea]{IRFU, CEA, Universit\'e Paris-Saclay, F-91191 Gif-sur-Yvette, France}
\address[miuni]{Dipartimento di Fisica, Universit\`a degli Studi di Milano, Via Celoria 16, I-20133 Milano, Italy}
\address[miinfn]{INFN, Sezione di Milano, Via Celoria 16, I-20133 Milano, Italy}
\address[surrey]{Department of Physics, University of Surrey, Guildford, GU2 7XH, United Kingdom}
\address[leuven]{KU Leuven, Department of Physics and Astronomy, Instituut voor Kern- en Stralingsfysica, B-3001 Leuven, Belgium}
\address[triumf]{TRIUMF, 4004 Wesbrook Mall, Vancouver, BC V6T 2A3, Canada}
\address[mcgill]{Department of Physics, McGill University, Montr\'eal, QC H3A 2T8, Canada}
\address[rcnp]{Research Center for Nuclear Physics (RCNP), Osaka University, Ibaraki 567-0047, Japan}
\address[ocu]{Department of Physics, Osaka City University, Osaka 558-8585, Japan}
\address[nitep]{Nambu Yoichiro Institute of Theoretical and Experimental Physics (NITEP), Osaka City University, Osaka 558-8585, Japan}
\address[cns]{Center for Nuclear Study, University of Tokyo, RIKEN campus, Wako, Saitama 351-0198, Japan}
\address[jaea]{Advanced Science Research Center, Japan Atomic Energy Agency, Tokai, Ibaraki 319-1195, Japan}
\address[caen]{LPC Caen, Normandie Univ, ENSICAEN, UNICAEN, CNRS/IN2P3, 14000 Caen, France}
\address[beijing]{State Key Laboratory of Nuclear Physics and Technology, Peking University, Beijing 100871, P.R. China}
\address[tohoku]{Department of Physics, Tohoku University, Sendai 980-8578, Japan}
\address[kth]{Department of Physics, Royal Institute of Technology, SE-10691 Stockholm, Sweden}
\address[ipno]{Universit\'e Paris-Saclay, CNRS/IN2P3, IJCLab, 91405 Orsay, France}
\address[lkb]{Laboratoire Kastler Brossel, Sorbonne Universit\'e, CNRS, ENS-PSL Research University, Coll\`ege de France, Case\ 74;\ 4, place Jussieu, F-75005 Paris, France}
\address[unal]{Universidad Nacional de Colombia, Carr. 30 No. 45-03, Bogot\'a, Colombia}
\address[rikkyo]{Department of Physics, Rikkyo University, 3-34-1 Nishi-Ikebukuro, Toshima, Tokyo 172-8501, Japan}
\address[inst]{Institute for Nuclear Science \& Technology, VINATOM, 179 Hoang Quoc Viet, Cau Giay, Hanoi, Vietnam}
\address[zagreb]{Ru{\dj}er Bo\v{s}kovi\'c Institute, Bijeni\v{c}ka cesta 54, 10000 Zagreb, Croatia}
\address[koeln]{Institut f\"ur Kernphysik, Universit\"at zu K\"oln, D-50937 Cologne, Germany}
\address[ewha]{Ewha Womans University, Seoul 120-750, South Korea}
\address[ibs]{Institute for Basic Science, Daejeon 34126, Korea}
\address[titech]{Department of Physics, Tokyo Institute of Technology, 2-12-1 O-Okayama, Meguro, Tokyo, 152-8551, Japan}
\address[hku]{Department of Physics, The University of Hong Kong, Pokfulam, Hong Kong}
\address[oslo]{Department of Physics, University of Oslo, N-0316 Oslo, Norway}
\address[atomki]{Institute for Nuclear Research, Atomki, P.O. Box 51, Debrecen H-4001, Hungary}
\address[lanzhou]{Institute of Modern Physics, Chinese Academy of Sciences, Lanzhou, China}

\begin{abstract}
  States in the $N=35$ and 37 isotopes \nuc{55,57}{Ca} have been populated by direct proton-induced nucleon removal reactions from \nuc{56,58}{Sc} and \nuc{56}{Ca} beams at the RIBF. In addition, the $(p,2p)$ quasi-free single-proton removal reaction from \nuc{56}{Ca} was studied. Excited states in \nuc{55}{K}, \nuc{55}{Ca}, and \nuc{57}{Ca} were established for the first time via in-beam $\gamma$-ray spectroscopy. Results for the proton and neutron removal reactions from \nuc{56}{Ca} to states in \nuc{55}{K} and \nuc{55}{Ca} for the level energies, excited state lifetimes, and exclusive cross sections agree well with state-of-the-art theoretical calculations using different approaches. The observation of a short-lived state in \nuc{57}{Ca} suggests a transition in the calcium isotopic chain from single-particle dominated states at $N=35$ to collective excitations at $N=37$.
\end{abstract}

\date{\today}
\end{frontmatter}

\paragraph*{Introduction}
The shell structure of atomic nuclei is one of the paradigms of modern physics. Established more than 70 years ago~\cite{mayer49,haxel49}, the concept of magic numbers as closed-shell configurations led to an easy understanding of the basic properties of many nuclei at or close to the valley of stability. Nuclei close to shell closures are dominated by single-particle configurations, while open shell nuclei far from magic numbers exhibit excitation spectra that are associated with collective and coherent motion of many nucleons.
Investigation of neutron-rich nuclei in the past decades, however, has shown that the magic numbers can change~\cite{otsuka20}. On the one hand, shell evolution leads to the absence of shell closures in, e.g., $N=8$~\cite{talmi60} or $N=20$~\cite{thibault75} nuclei. On the other hand, for isotopes with closed-proton configurations, new neutron magic numbers, not observed for nuclei close to stability, were experimentally established in neutron-rich nuclei. For instance, the isotope \nuc{24}{O} was identified as doubly-magic nucleus through measurements of separation energies~\cite{hoffman08} and the high excitation energy for the first $2^+$ state at 4.7~MeV~\cite{hoffman09}, reflecting a new $N=16$ shell gap between the $\nu1s_{1/2}$ and $0d_{3/2}$ orbitals. In addition, spectroscopic factors for the neutron knockout from \nuc{24}{O} were interpreted in terms of a dominant $(1s_{1/2})^2$ ground-state configuration~\cite{kanungo09}, consistent with the expectation that all levels below the $N=16$ sub-shell closure are fully occupied.
Similarly, along the $Z=20$ Ca isotopic chain, new sub-shell closures at $N=32$ and $34$ were first inferred from the excitation energies of the $2^+_1$ states~\cite{huck85,steppenbeck13}, pointing to the magic nature of these isotopes. The two-neutron separation energies extracted from the measured masses~\cite{wienholtz13,michimasa18} support the picture of large energy gaps between the neutron $1p_{3/2}$, $1p_{1/2}$, and $0f_{5/2}$ orbitals.
Recently, the analysis of one-neutron knockout reactions populating states in \nuc{53}{Ca} further supported the magicity of \nuc{54}{Ca}~\cite{chen19}. In particular, the shell model picture describing the ground state of \nuc{54}{Ca} with fully occupied neutron $1p_{1/2}$ and $1p_{3/2}$ orbitals and an empty $0f_{5/2}$ orbital was corroborated by these findings.

In this picture, the single-particle structure of the K isotopes is governed by the protons in the $sd$ shell and the expected ground state spin-parity is $3/2^+$ for a single proton hole at $Z=19$. In neutron-rich isotopes of potassium, a narrowing of the gap between the $\pi s_{1/2}$ and the $\pi d_{3/2}$ orbital occurs. The ground state of \nuc{47}{K}, at $N=28$, has $J^\pi = 1/2^+$, while the $L=2$ strength measured in electron-induced knockout reactions is concentrated in a single excited $3/2^+$ state~\cite{kramer01}. Laser spectroscopy measurements found a spin-parity of $1/2^+$ also for the ground state of \nuc{49}{K}, while the normal ordering of levels is restored in \nuc{51}{K}~\cite{papuga13}. In agreement with theoretical calculations using a variety of many-body methods and interactions, the proton knockout reactions from \nuc{52,54}{Ca} strongly populate the $3/2^+$ ground and $1/2^+$ excited states in the $N=32,34$ isotopes \nuc{51,53}{K}~\cite{sun20}. 

It is therefore interesting to study the evolution of single-particle states further along the Ca isotopic chain. The proton shell closure isolates the degrees of freedom and allows us to probe both the proton structure for isotopes beyond $N=34$ and to get first experimental information on the neutron states in the transitional region between the $N=34$ closed shell and a proposed island of inversion for neutron-rich $N=40$ isotones which is characterized by deformed ground states and low-lying collective excitations~\cite{lenzi10}. 

In this letter, we report on the first spectroscopy of \nuc{55,57}{Ca} and \nuc{55}{K}, which are the first beyond the $N=34$ shell closure to be studied.
Since the spectroscopic information of odd-mass nuclei is beneficial for elucidating the single-particle structure, we present the experimental result of $^{55}$K for understanding the proton shell structure, and the results of $^{55}$Ca and $^{57}$Ca for the neutron shell structure.
The present measurements reveal that the $Z=20$ proton shell closure remains intact beyond $N=34$. First spectroscopic information on states in \nuc{55,57}{Ca}, combined with a theoretical analysis, suggest a transition from a simple particle-hole configuration for the first excited state in \nuc{55}{Ca} to a more collective excitation in \nuc{57}{Ca}.

\paragraph*{Experimental setup and results}
The experiment was performed at the Radioactive Isotope Beam Factory, operated by RIKEN Nishina Center and Center for Nuclear Study, University of Tokyo. Very neutron-rich nuclei were produced by fragmenting a 345~$A$MeV \nuc{70}{Zn} beam on a 10~mm thick Be primary target. The reaction products were selected and identified in the BigRIPS separator~\cite{kubo12} using the TOF$-\Delta E-B\rho$ method~\cite{fukuda13}.
The intensity of the secondary \nuc{56}{Ca} beam was 0.16 particles per second and constituted $8.6\cdot 10^{-4}$ of the total secondary beam intensity.
 Its energy before the target amounted to 250~$A$MeV with a 3\% momentum spread.
The beam impinged on the $151(1)$~mm thick liquid hydrogen target system MINOS~\cite{obertelli14}. The target was surrounded by a 300~mm long time-projection chamber (TPC) to measure the trajectories of recoiling protons and protons removed from the projectiles. The tracking of the beam particles, facilitated by multi-wire drift chambers before the target, and the protons allowed the reconstruction of the reaction vertex~\cite{santamaria18} and therefore the velocity and $\gamma$-ray emission angle used for Doppler correction. The efficiency of the MINOS TPC and the proton tracking reconstruction for detecting at least one proton amounted to $92(5)$ and $71(5)$\% for the \pppf{56}{Ca}{55}{K} and \ppnf{56}{Ca}{55}{Ca} reactions, respectively.
Reaction products were analyzed by the SAMURAI spectrometer~\cite{kobayashi13}. The large gap magnet was operated at 2.7~T. Event-by-event identification was achieved by measuring the trajectory before and after the magnet with two sets of drift chambers as well as the kinetic energy and time-of-flight in a plastic scintillator array placed after the SAMURAI dipole magnet.

$\gamma$ rays emitted from excited states in the final nuclei were detected with the upgraded DALI2$^+$ array~\cite{takeuchi14}. Its 226 NaI(Tl) scintillator detectors were calibrated using standard calibration sources. The in-beam energy resolution amounted to 11\% (FWHM) for $\approx 500-700$~keV, the energy range of interest of this work. The full energy peak efficiency after add-back of hits in crystals within a 15~cm radius was 28\% at 1.3~MeV.
The response of the array to \grays emitted at $v\approx 0.6 c$ was simulated using the GEANT4 toolkit~\cite{agostinelli03}. Experimental \ghray energy spectra were then fitted with the response functions with varying transition energies and lifetimes in addition to a double exponential background.

\begin{figure}[h]
\centering
\includegraphics[width = \columnwidth]{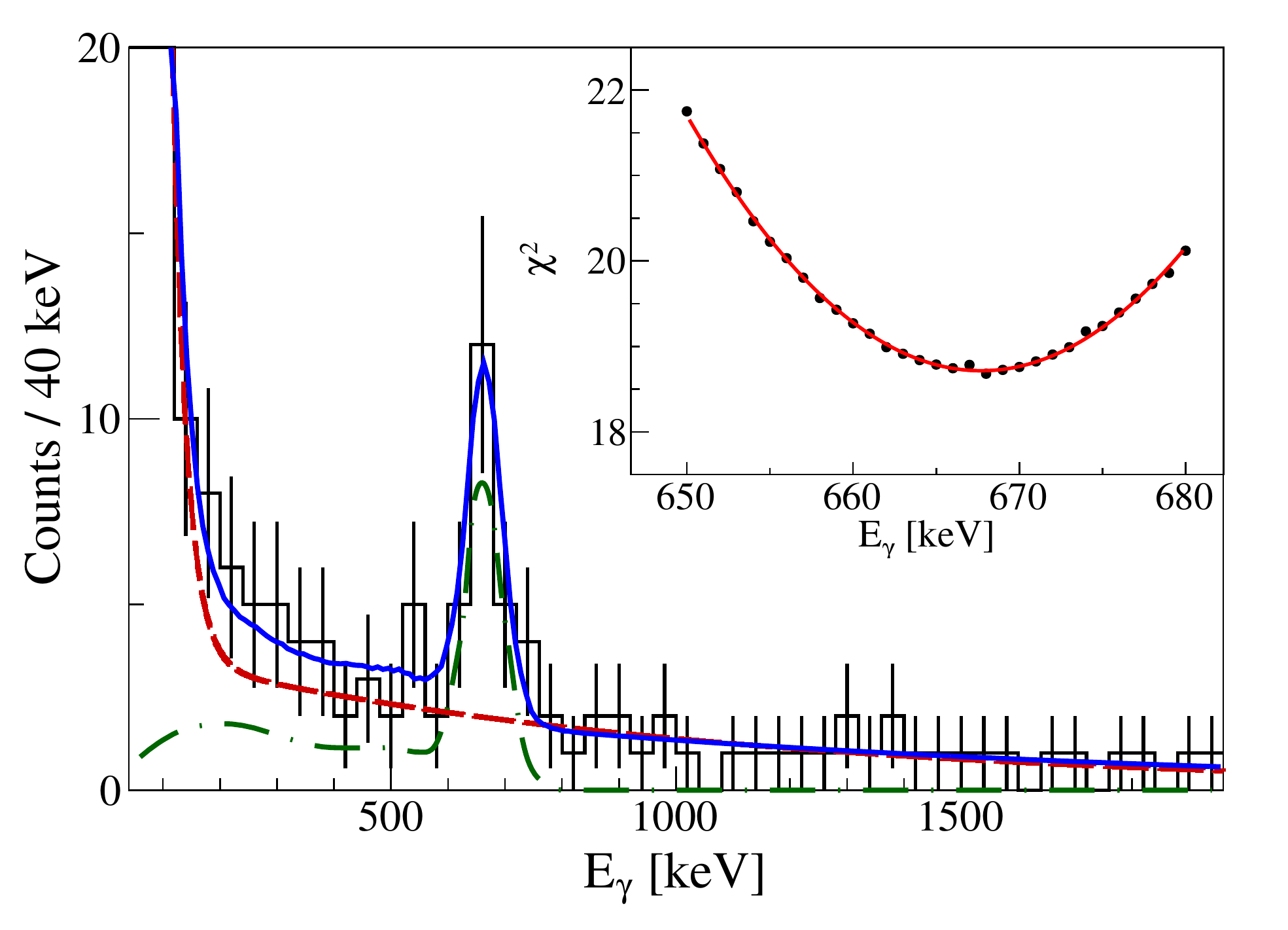}
\caption{Doppler corrected \ghray energy spectrum for the \pppfr{56}{Ca}{55}{K} one-proton knockout reaction. The data (black) are fitted with a simulation (blue, solid) composed of the simulated response function (green, dotted-dashed) and a double exponential background (red, dashed). The inset shows the $\chi^2$ distribution as a function of the assumed transition energy. The lifetime was set to 0~ps in this case.}
\label{fig:spectrum55k}
\end{figure}

An inclusive cross section of $5.3(5)$~mb was measured for the \pppf{56}{Ca}{55}{K} one-proton knockout reaction to bound final states. For the uncertainty determination, the acceptance of the SAMURAI spectrometer, the losses of beam and reaction products in the target, and the efficiency of the beam line detectors were considered.
Fig.~\ref{fig:spectrum55k} shows the Doppler corrected \ghray energy spectrum measured for the \pppfr{56}{Ca}{55}{K}.
A single transition at 668(10)~keV was observed, where the uncertainty includes statistical uncertainties from the fit as well as systematic sources from the energy calibration and the MINOS vertex reconstruction uncertainty. In the fit, a negligible lifetime has been assumed and an upper limit of $\tau < 53$~ps can be extracted from the peak shape. By varying the assumed lifetime within these limits, one obtains an additional systematic uncertainty for the extracted energy of less than $10$~keV. The exclusive cross sections for the ground and 668~keV states are shown in Table~\ref{tab:res}, assuming that this transition corresponds to the de-excitation of a single bound excited state. 
\begin{table*}[h]                 
  \caption{Transition energies, cross sections, and spectroscopic factors for states populated in \nuc{55}{K}, \nuc{55}{Ca}, and \nuc{57}{Ca}. Spin and parity assignments are based on systematics and theoretical calculations. Spectroscopic factors were extracted by dividing the measured cross sections by the calculated single-particle ones averaged for the beam energy at the reaction vertex. Experimental results for the energies and cross sections are displayed together with theoretical calculations using several approaches, for details see text.}
  \begin{center}
    \label{tab:res}
    \begin{threeparttable}
      {\renewcommand{\arraystretch}{1.2}
        \begin{tabular}{rrrcrrrrrrrrr} 
          \hline
          \multicolumn{3}{c}{Experiment}      &         &                         & \mcnt{LSSM}         & \mcnt{VS-IMSRG}     & \mcnt{SCGF NNLO$_\text{sat}$} & \mcnt{SCGF NN+3N(lnl)}  \\
          $E$ (keV) & $\sigma$ (mb) & $C^2S$   & $J^\pi$ & $\sigma_\text{sp}$      & $E$ (keV) & $C^2S$ & $E$ (keV) & $C^2S$ & $E$ (keV) & $C^2S$      & $E$ (keV) & $C^2S$ \\    
          \hline                                                                   
          \multicolumn{12}{c}{ \pppf{56}{Ca}{55}{K} } \\                           
          \hline                                                                   
          0         & 3.5(7)        & 3.09(62) & $3/2^+$ & 1.15                    &    0      & 3.08   &    0      & 2.88   &  252      & 2.35        &   0       & 3.02   \\
          668(10)   & 1.7(5)        & 1.31(42) & $1/2^+$ & 1.30                    &  451      & 1.06   &  200      & 1.24   &    0      & 1.25        & 966       & 1.52   \\
          \hline
          \multicolumn{12}{c}{ \ppnf{56}{Ca}{55}{Ca} } \\
          \hline
          0         &  $8.3^{+6.7}_{-3.6}$ & $2.0^{+1.6}_{-0.9}$ & $5/2^-$ & 4.17 &    0     & 2.12   &    0      & 2.09   &    0      & 1.54        &    0      & 1.65   \\
          673(17)   & $19.6^{+6.7}_{-3.7}$ & $3.0^{+1.0}_{-0.6}$ & $1/2^-$ & 6.52 &  864     & 1.78   & 1352      & 1.84   &  241      & 1.52        & 1096      & 1.74   \\
                    &               &                            & $3/2^-$ &      & 3152     & 3.30   & 3918      & 3.68   & 2594      & 2.87        & 3513      & 3.29   \\
          \hline
          \multicolumn{12}{c}{ \pppf{58}{Sc}{57}{Ca} } \\
          \hline
          0         & 0.75(7)       &          & $5/2^-$ &                        &    0      & 0.68\fn{1}   &    0      & 0.61\fn{2}   &   0       &             &    0      &        \\
          751(13)   & 0.18(3)       &          & $3/2^-$ &                        &  653      & 0.20\fn{1}   &  495      &              &           &             &           &        \\
                    &               &          & $1/2^-$ &                        &  816      &              & 1595      &              &  325         &             &  1413      &        \\
                    &               &          & $9/2^-$ &                        & 1206      & 0.10\fn{1}   &  983      &              &           &             &           &        \\       
          \hline
        \end{tabular}
      }
      \begin{tablenotes}
      \item[1]{The predicted $5^+$ ground state is used for \nuc{58}{Sc}.}
      \item[2]{The predicted $2^+$ ground state is used for \nuc{58}{Sc}.}
      \end{tablenotes}
    \end{threeparttable}
  \end{center}
\end{table*}
Parallel momentum distributions of the \nuc{55}{K} ejectiles could not be constructed due to limited statistics.

\begin{figure}[h]
\centering
\includegraphics[width = \columnwidth]{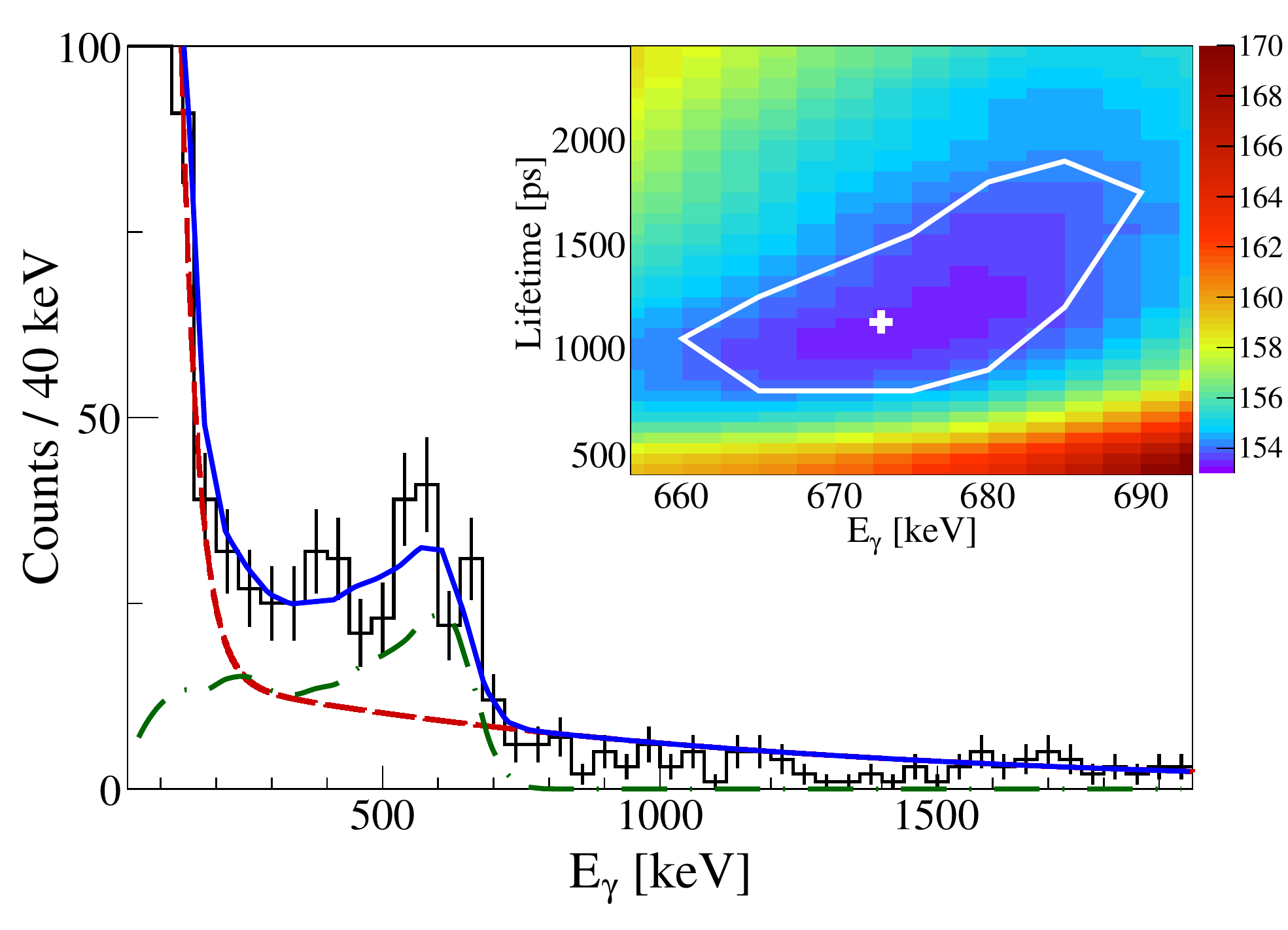} 
\caption{Same as Fig.~\ref{fig:spectrum55k}, but for the \ppnfr{56}{Ca}{55}{Ca}. The inset shows the two-dimensional $\chi^2$ distribution as a function of the assumed transition energy and the lifetime of the state. The cross indicates the energy and lifetime that minimize $\chi^2$ while the white line is the 1-$\sigma$ contour line.}
\label{fig:spectrum55ca}
\end{figure}

For the one-neutron knockout reaction from \nuc{56}{Ca}, the inclusive reaction cross section amounted to 27.9(23)~mb. In the Doppler corrected \ghray energy spectrum, shown in Fig.~\ref{fig:spectrum55ca}, a broad structure is observed instead of a sharp peak.
This is consistent with the expectation of an unhindered single-particle $E2$ transition for the decay to the ground state.
For example, the $E2$ decay of a state at 673~keV with a reduced transition probability of 1~W.u. would result in a level lifetime of 476~ps. The mean decay position was thus located 107~mm after the reaction vertex, leading to a low energy tail in the Doppler corrected \ghray energy spectrum as observed in Fig.~\ref{fig:spectrum55ca}. A combined fit of both transition energy and level lifetime results in best fit values of $E = 673(17)$~keV and $\tau = 1130^{+520}_{-330}$~ps. The parameters for the double exponential background included in the fit were constrained from fits to spectra of neighboring nuclei for analogue reactions. Alternatively, the spectrum can be fitted with two or more transitions with negligible lifetimes. However, a coincidence can be ruled out and the occurrence of two low-lying states populated in the reaction is at variance with the expectation of the ground state configuration of \nuc{56}{Ca} and theoretical calculations discussed below. 
It should be noted that the extracted intensity of the $673(17)$~keV transition depends strongly on the lifetime assumed in the simulation of the response function. For the extraction of the exclusive cross sections shown in Table~\ref{tab:res} this correlation was taken into account. 

\begin{figure}[h]
\centering
\includegraphics[width = \columnwidth]{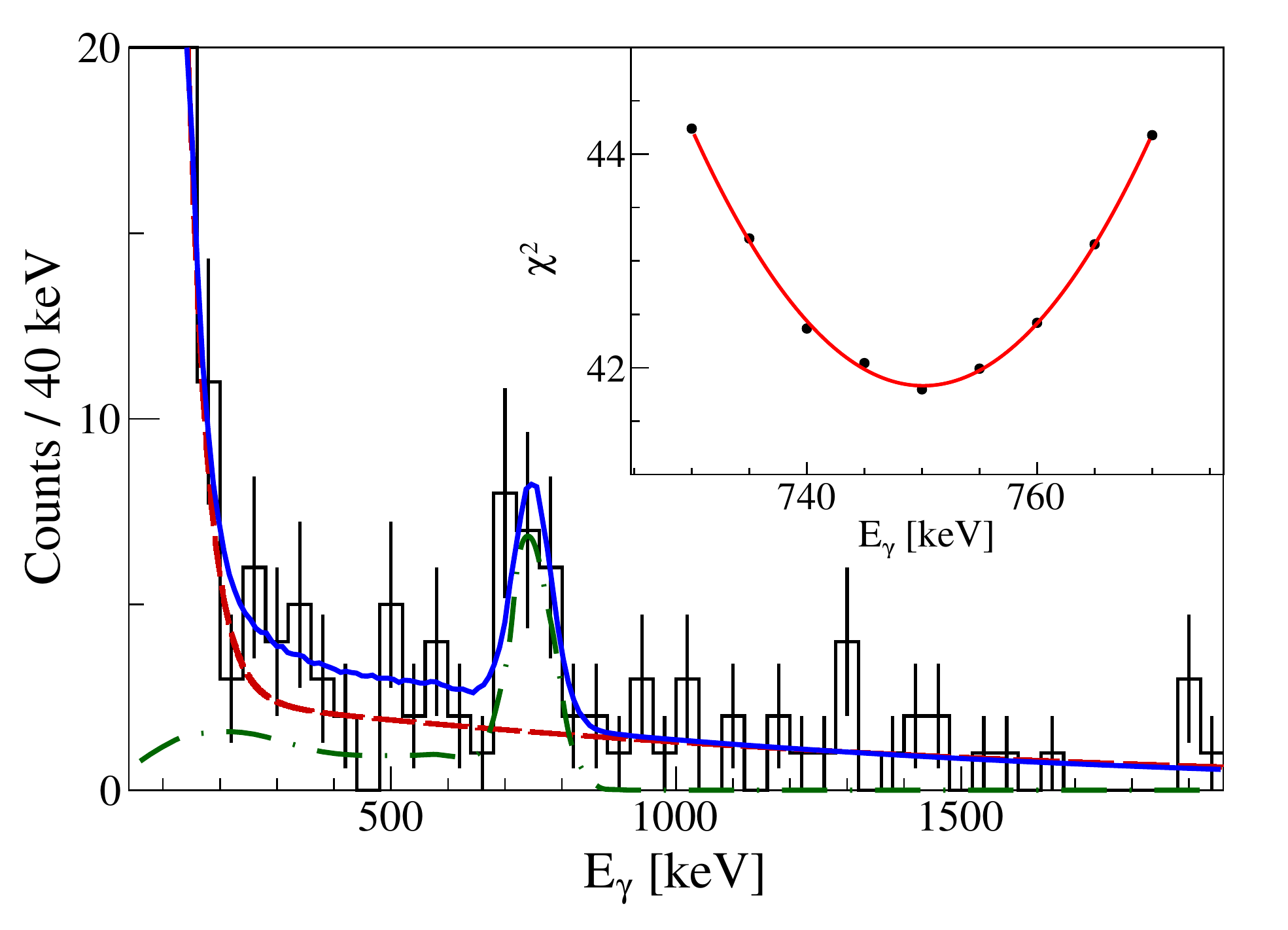}
\caption{Same as Fig.~\ref{fig:spectrum55k}, but for the \pppfr{58}{Sc}{57}{Ca}. The inset shows the $\chi^2$ distribution as a function of the assumed transition energy of the decay.}
\label{fig:spectrum57ca}
\end{figure}

States in \nuc{57}{Ca} were populated in the proton removal from \nuc{58}{Sc}. In the same setting as employed for the \nuc{56}{Ca} study, the \nuc{58}{Sc} beam intensity amounted to 1.2~pps (purity of 0.65\%).
The Doppler corrected \ghray energy spectrum for the \pppfr{58}{Sc}{57}{Ca} is shown in Fig.~\ref{fig:spectrum57ca}.
A single transition at $751(13)$~keV with a statistical significance of $6.1\sigma$ above background was observed. The short lifetime, $\tau < 62$~ps, originates from a fast $M1$ or a collective $E2$ transition.

Experimental spectroscopic factors were obtained by dividing the experimental exclusive cross section by calculated single-particle cross sections. These employ the distorted wave impulse approximation~\cite{wakasa17} and were already applied to the proton and neutron removal from \nuc{54}{Ca}~\cite{sun20,chen19}.
Single-particle wave functions and the nuclear density were obtained by the Bohr-Mottelson single-particle potential~\cite{bohr69}. The depth parameter is adjusted to reproduce the experimental separation energies~\cite{michimasa18,michimasa20,wang17}. The optical potential is constructed by the folding model~\cite{toyokawa13} with the Melbourne g-matrix interaction~\cite{amos00} and the calculated nuclear density. The spin-orbit term of the optical potential is omitted in the present calculations. The Franey-Love effective interaction~\cite{franey85} is adopted for the $pN$ collision.

\paragraph*{Discussion}
In a naive picture, the ground state of \nuc{56}{Ca} is composed of two neutrons in the $0f_{5/2}$ orbital on top of a closed \nuc{54}{Ca} core.
Proton removal from \nuc{56}{Ca} can occur from the $\pi 1s_{1/2}$ or $0d_{3/2}$ orbitals populating $1/2^+$ and $3/2^+$ hole states in \nuc{55}{K}.
In contrast to the work presented in Ref.~\cite{sun20}, parallel momentum distributions of the \nuc{55}{K} ejectiles could not be analyzed in the present work due to sparse statistics.
The spectroscopic factors presented in Table~\ref{tab:res} are close to the independent particle limit $C^2S = 2j+1$ for a fully occupied $j$ orbital when taking into account the typical quenching factor of $\sim 0.6$~\cite{aumann21}. 
This suggests a $(3/2^+)$ assignment for the ground state of \nuc{55}{K} and $(1/2^+)$ for the excited state at 668~keV. 

For a more rigorous analysis, the experimental results are confronted with several nuclear structure calculations as shown in Table~\ref{tab:res}. The large scale shell model (LSSM) calculations employ the SDPF-MUs effective interaction for the proton $sd$-shell and neutron $pf$-shell model space. 
This interaction is a modification of the SDPF-MU Hamiltonian~\cite{utsuno12} combined with the GXPF1Bs interaction~\cite{chen19} to better describe the data of the neutron-rich K isotopes~\cite{papuga13,sun20}. 
The ab initio valence-space in-medium similarity renormalization group (VS-IMSRG)~\cite{stroberg19} calculations were done employing the 1.8/2.0 (EM) interaction~\cite{hebeler11,simonis17} which well reproduces ground-state energies to the heavy-mass region~\cite{stroberg21,morris18}.
Starting in a harmonic-oscillator basis of 15 major-shells, with an oscillator energy of $16$~MeV, the Hamiltonian is transformed to the Hartree-Fock reference state with the ensemble normal ordering to best capture the effect of the three-nucleon interaction acting on valence particles~\cite{stroberg17}.
An additional boundary for the three-nucleon interaction $E_{\rm 3max}=24$~\cite{miyagi21} is introduced as the sum of the three-body harmonic oscillator quantum numbers in the present calculations.
Using an approximate unitary transformation from the Magnus formalism~\cite{morris15}, the effective Hamiltonian for the proton $sd$ and neutron $fp$ space above a \nuc{28}{O} core was calculated. All the operators are truncated to the two-body level during the VS-IMSRG calculation done with the \textbf{imsrg++} code~\cite{stroberg}. In both the LSSM and VS-IMSRG calculations, the diagonalization in the valence space was performed with the KSHELL code~\cite{shimizu19}.
Lastly, full-space self-consistent Green's function (SCGF)~\cite{soma20b} calculations were performed in the Gorkov ADC(2) approximation scheme~\cite{soma11, soma14} using the BcDor code~\cite{barbieri}.
Two different two- plus three-body ($NN+3N$) interactions derived within chiral effective field theory were used, NNLO$_\text{sat}$~\cite{ekstroem15} and $NN+3N$(lnl)~\cite{soma20}. 
SCGF calculations employed a 14 major-shell harmonic oscillator basis with an additional truncation on $3N$ matrix elements at $E_{\rm 3max}=16$.

All theoretical results agree in predicting a $3/2^+$ ground state for \nuc{55}{K}, except for the calculations with the NNLO$_\text{sat}$ interaction that suggest a $1/2^+$ ground state. 
This disagreement has been already observed for lighter potassium isotopes, with NNLO$_\text{sat}$ consistently producing a too strongly bound $1/2^+$ state along the whole chain~\cite{sun20}.
The other three calculations yield an excited $1/2^+$ state whose energy is similar to the experimentally observed value, with VS-IMSRG and LSSM (SCGF) slightly underestimating (overestimating) it.
Furthermore, within each of the four theoretical schemes, spectroscopic factors for the \pppf{56}{Ca}{55}{K} reaction point to a substantial occupancy of both the $0d_{3/2}$ and $1s_{1/2}$ orbitals.
This is again consistent with the expectation of a closed $Z=20$ shell which remains intact also at $N=36$. 
In view of these results, we tentatively assign spin-parities of $(3/2^+)$ to the ground and $(1/2^+)$ to the 668~keV state in \nuc{55}{K}.

Moving to \nuc{55}{Ca}, let us note that the single-particle character of the $673(17)$~keV state is evident from its large population in the one-neutron knockout reaction.  
The one-neutron separation energy of \nuc{55}{Ca} amounts to only 1561(167)~keV~\cite{michimasa18}, therefore neutron knockout from the more deeply bound $1p_{3/2}$ orbital likely leads to the population of unbound states. The population of two states in \nuc{55}{Ca} with roughly equal spectroscopic strength is consistent with the expectation that the ground state of \nuc{55}{Ca} contains a single neutron in the $0f_{5/2}$ orbital, while the excited state results from the promotion of a neutron from the $1p_{1/2}$ orbital across the $N=34$ sub-shell closure. Such a $1p-1h$ state is easily populated in the neutron knockout from \nuc{56}{Ca} for which a $(0f_{5/2})^2$ configuration is expected.
The proton removal reaction from the \nuc{56}{Sc} isotone was also analyzed, but no evidence for the population of bound excited states was found. The removal of the $0f_{7/2}$ valence proton from the $(1^+)$ ground state of \nuc{56}{Sc}~\cite{liddick04} leads to the population of the \nuc{55}{Ca} ground state.

The experimental results for the \ppnf{56}{Ca}{55}{Ca} reaction are in good agreement with large-scale shell model calculations using the GXPF1Bs effective interaction~\cite{chen19}.
The interaction is based on the standard GXPF1B interaction~\cite{honma08} with an adjustment of the $\nu p_{3/2}-\nu f_{5/2}$ monopole strength~\cite{steppenbeck13} and the $(\nu f_{7/2})^2$ and $(\nu f_{5/2})^2$ pairing matrix elements for better description of neutron-rich Ca isotopes~\cite{chen19}. Note that a calculation with the original GXPF1B effective interaction \cite{honma08} leads to a much larger excitation energy of the first $1/2^-$ state, while calculations with the KB3G interaction~\cite{poves01}, which does not reproduce the $N=34$ shell closure, give a much smaller excitation energy of only 225~keV.
The other theoretical approaches also agree in the level ordering with a $5/2^-$ ground state and a $1/2^-$ excited state (see Table~\ref{tab:res}). The excitation energy predicted by VS-IMSRG calculations is too high, suggesting a too strong $N=34$ shell closure, while in the SCGF calculations with  NNLO$_\text{sat}$ a much smaller energy is predicted.
The ab initio interaction that better describes the measured $1/2^-$ energy is $NN+3N$(lnl), with a value slightly above the shell-model result.
All calculations predict a $3/2^-$ state with significant spectroscopic factor, but in this experiment such a state is inaccessible due to its location above the neutron separation energy. 

Spectroscopic factors computed within each of the theoretical approaches are presented in Table~\ref{tab:res}. Each calculation leads to interpreting the first excited state in \nuc{55}{Ca} as a dominant neutron $(1p_{1/2})^{-1}(0f_{5/2})^2$ configuration. Another experimental hint at the neutron dominance of the 673~keV state in \nuc{55}{Ca} comes from the fact that this state is not observed in the proton removal from \nuc{56}{Sc}.
The long lifetime corresponds to a $B(E2;\; (1/2^-) \rightarrow (5/2^-)) = 5.2$~\efm in reasonable agreement with the LSSM calculations which predict 2 \efm. The VS-IMSRG calculations predict a value $B(E2;\; 1/2^- \rightarrow 5/2^-) = 0.6$~\efm but are found in general to under-predict $E2$ matrix elements compared to experimental results.

In contrast to \nuc{55}{Ca}, the excited state observed in \nuc{57}{Ca} is populated in the proton removal reaction from its isotone \nuc{58}{Sc}. The latter nucleus has been discussed as a member of the island of inversion centered around \nuc{64}{Cr} and shell model calculations with the LNPS effective interaction~\cite{lenzi10} show that the wave functions of low-lying states contain significant contributions from particle-hole configurations~\cite{wimmer21}. The quasi-free $(p,2p)$ reaction on such a system is likely to leave the neutron configuration untouched and, therefore, will populate states with neutron particle-hole excitations.
This suggests that the newly observed state at 751~keV is of collective rather than single-particle nature. The fast decay with a lifetime of less than 62~ps corroborates this interpretation. Assuming an $E2$ decay, the present upper limit corresponds to a lower limit for the $B(E2)$ value of 55.2~\efm
, much more collective than the value determined for the decay of the single-particle state in \nuc{55}{Ca}.

Since the ground state spin and parity of \nuc{58}{Sc} are not known and this nucleus furthermore has at least one long-lived isomeric state~\cite{wimmer21}, meaningful spectroscopic factors cannot be extracted from the experimental cross section. The LSSM and VS-IMSRG calculations are able to predict spectroscopic factors. In the former calculations, large spectroscopic factors for the removal of a $0f_{7/2}$ proton are found for the $5/2^-$ ground and $3/2^-$ first excited state, rather independently of the assumption of the \nuc{58}{Sc} ground state. For the VS-IMSRG calculations, the model space was changed with the prescription introduced in Ref.~\cite{miyagi21} to include the $\pi 0f_{7/2}$ orbital at the expense of the $0d_{5/2}$ one for a \nuc{34}{Si} core. In this calculation, the first excited state of \nuc{57}{Ca} is predicted as $3/2^-$ with a $\nu (0f_{5/2})^3$ dominated configuration. 
Full-space SCGF calculations contain less collective degrees of freedom compared to valence-space approaches.
As a consequence, the first excited $3/2^-$ state is out of reach. 
An excited $1/2^-$ level, of single-particle character, is instead found around 1.5~MeV for the $NN+3N$(lnl) interaction, in agreement with VS-IMSRG results.
Nevertheless, a careful inspection of the SCGF spectral distribution reveals a significantly more fragmented strength in the $3/2^+$ and $1/2^+$ proton removal channels from \nuc{58}{Ca} with respect to \nuc{56}{Ca}, signaling the appearance of more collective correlations.
This is also in line with a worsening of the description of binding energies along the Ca chain past $N=36$ discussed in Ref.~\cite{soma21}.
All these findings, together with the increase in separation energies~\cite{michimasa18} for \nuc{57}{Ca}, suggest a transition towards more deformed ground states beyond $N=36$.

\paragraph*{Conclusions}
In summary, we performed first spectroscopic studies of the neutron-rich \nuc{55}{K} and \nuc{55,57}{Ca} isotopes by means of direct nucleon knockout reactions. The first excited state at 668(10)~keV in \nuc{55}{K} is tentatively assigned a spin-parity of $(1/2^+)$ with a $(3/2^+)$ ground state based on the measured cross sections and theoretical calculations.
This observation is in agreement with the rigid closed proton $Z=20$ core observed in the calcium chain and complements the data obtained for the lighter K isotopes.
The neutron removal reaction from \nuc{56}{Ca} is consistent with the interpretation of a simple $\nu (0f_{5/2})^2$ ground state configuration leading to the population of the $(5/2^-)$ ground state and $(1/2^-)$ first excited state at 673(17)~keV in \nuc{55}{Ca}. The reduced transition probability for the decay of the $1/2^-$ state extracted from the measured lifetime is compatible with a single-particle state as well.
Finally, the population of a short-lived state in \nuc{57}{Ca} at 751(13)~keV suggests a structural change beyond $N=36$ and a transition toward many particle-hole dominated configurations in the $N=40$ Island of Inversion. In the future, more detailed experimental work on the very neutron-rich Ca isotopes will allow to further characterize the interplay between single-particle and collective excitations in this region and benchmark the different theoretical approaches.

\paragraph*{Acknowledgments}
We would like to thank the RIKEN accelerator and BigRIPS teams for providing the high intensity beams.
T.K.\ acknowledges support by RIKEN Junior Research Associate Program.
K.W.\ acknowledges support from the Spanish Ministerio de Ciencia, Innovaci\'on y Universidades RYC-2017-22007.
J.D.H. and T.M. would like to thank S.~R.~Stroberg for the imsrg++ code~\cite{stroberg} used to perform VS-IMSRG calculations.
TRIUMF receives funding via a contribution through the National Research Council of Canada.
J.D.H is further supported by NSERC under grants SAPIN-2018-00027 and RGPAS-2018-522453.
VS-IMSRG computations were performed with an allocation of computing resources on Cedar at WestGrid and Compute Canada, and on the Oak Cluster at TRIUMF managed by the University of British Columbia department of Advanced Research Computing (ARC).
N.S. and Y.U. acknowledge valuable support by "Priority Issue on post-K computer" and KAKENHI grant 20K03981 and 17K05433.
C.B.  was supported by the UK Science and Technology Facilities Council (STFC) through grants No. ST/L005816/1 and No. ST/V001108/1.
SCGF calculations were performed by using HPC resources from GENCI-TGCC, France (Contract No. A009057392) and  at the DiRAC DiAL system at the University of Leicester, UK, (funded  by  the  UK  BEIS via  STFC  Capital  Grants No.  ST/K000373/1 and No.  ST/R002363/1 and STFC DiRAC  Operations  Grant  No.  ST/R001014/1)..
I.M. was supported by the RIKEN IPA program, F.B. by the RIKEN Special Postdoctoral Researcher Program.
D.S. acknowledges support from the European Regional Development Fund contract No. GINOP-2.3.3-15-2016-00034 and the National Research, Development and Innovation Fund of Hungary via Project No. K128947.
K. I. H., D. K., and S. Y. P. acknowledge the support from the IBS grant funded by the Korea government (No. IBS-R031-D1).
The work was further supported by JSPS KAKENHI Grant Nos. JP16H02179, JP18H05404, JP19H00679, and JP21H01114 and the Deutsche Forschungsgemeinschaft (DFG) under Grant No. BL 1513/1-1.
\bibliographystyle{elsarticle-num-names}
\bibliography{draft}

\end{document}